\documentclass[12pt,a4paper,fleqn]{article}

\usepackage[textheight=23cm,textwidth=15cm]{geometry}
\usepackage{amsmath}
\usepackage{upgreek}
\usepackage{graphicx}
\usepackage{multirow}
\usepackage[american]{babel}
\usepackage{here}
\usepackage[articletitle=true, maxauthors=30]{achemso}
\usepackage{xcolor}
\usepackage{siunitx}
\usepackage{subcaption}

\begin{document}

\begin{center}

{\LARGE\bf
Hierarchical quantum embedding by machine learning
for large molecular assemblies
}

\vspace{0.5cm}

{\large
Moritz Bensberg$^{a}$, Marco Eckhoff$^{a}$, Raphael T. Husistein$^{a}$, Matthew S. Teynor$^{b,c}$, Valentina Sora$^{d}$, William Bro-J{\o}rgensen$^{b}$, F. Emil Thomasen$^{e}$, Anders Krogh$^{d,f}$\footnote{email: akrogh@di.ku.dk}, Kresten Lindorff-Larsen$^e$\footnote{email: lindorff@bio.ku.dk}, Gemma C. Solomon$^{b,c}$\footnote{email: gsolomon@chem.ku.dk}, Thomas Weymuth$^a$, and Markus Reiher$^{a}$
\footnote{email: mreiher@ethz.ch}
}\\[4ex]

$^{a}$ ETH Zurich, Department of Chemistry and Applied Biosciences, Vladimir-Prelog-Weg 2, 8093 Zurich, Switzerland

$^{b}$ University of Copenhagen, Department of Chemistry and Nano-Science Center, Universitetsparken 5, DK-2100, Copenhagen \O, Denmark

$^{c}$ University of Copenhagen, Niels Bohr Institute, NNF Quantum Computing Programme, Blegdamsvej 17, DK-2100 Copenhagen \O, Denmark

$^{d}$ University of Copenhagen, Department of Computer Science, Universitetsparken 1, DK-2100, Copenhagen \O, Denmark

$^{e}$ University of Copenhagen, Department of Biology, Linderstr{\o}m-Lang Centre for Protein Science, Ole Maal{\o}es Vej 5, DK-2200, Copenhagen N, Denmark

$^{f}$ University of Copenhagen, Center for Health Data Science, Department of Public Health, {\O}ster Farimagsgade 5, DK-1353, Copenhagen \O, Denmark

March 5, 2025

\vspace{0.3cm}

\textbf{Abstract}
\end{center}
\vspace*{-0.6cm}
{\footnotesize{
We present a quantum-in-quantum embedding strategy coupled to machine learning potentials to improve on the accuracy of quantum-classical hybrid models for the description of large molecules. In such hybrid models, relevant structural regions (such as those around reaction centers or pockets for binding of host molecules) can be described by a quantum model that is then embedded into a classical molecular-mechanics environment. However, this quantum region may become so large that only approximate electronic structure models are applicable. To then restore accuracy in the quantum description, we here introduce the concept of quantum cores within the quantum region that are amenable to accurate electronic structure models due to their limited size. Huzinaga-type projection-based embedding, for example, can deliver accurate electronic energies obtained with advanced electronic structure methods. The resulting total electronic energies are then fed into a transfer learning approach that efficiently exploits the higher-accuracy data to improve on a machine learning potential obtained for the original quantum-classical hybrid approach. We explore the potential of this approach in the context of a well-studied protein-ligand complex for which we calculate the free energy of binding using alchemical free energy and non-equilibrium switching simulations.
}}
\vspace{-0.7cm}

\section{Introduction}
\label{sec:intro}
Hybrid models for large macromolecules can leverage a structural decomposition into a scaffold (spectator) part, which can be described by classical molecular mechanics (MM) force fields, and into a smaller quantum region described by a quantum mechanical (QM) method. Such a decomposition is routinely applied to describe local phenomena, such as catalytic chemical reactions mediated by metal ions in metalloenzymes or the binding of drug molecules to proteins. However, to obtain reliable results that do not suffer from artifacts caused by the MM description of the scaffold, large QM regions are required\cite{Sumowski2009, Hu2011, Liao2013, Kulik2016} to which only approximate electronic structure methods (such as density functional theory (DFT) or even fully semi-empirical quantum chemical approaches) can be applied. More accurate electronic structure models are hardly applicable because of their steep scaling of computational costs with system size.
The accuracy of DFT for the QM region may be enhanced by quantum-in-quantum (QM/QM) embedding approaches\cite{Knizia2013a, Jacob2014, Jacob2024, Libisch2014, Welborn2016, Lee2019}. 

Such QM/QM/MM approaches were used previously to study the reactivity in protein complexes (see Refs.~\citenum{Bennie2016, Zhang2018, Hegely2018, Macetti2021a} for examples). However, more general biochemical problems such as the calculation of binding free energies have not been targeted with modern DFT-based QM/QM embedding approaches. Here, we describe such a general approach, where we exploit a machine learning potential (MLP) representation to mediate between the different types of data to be combined in a single hybrid model. First, we rely on our previous work \cite{Q4Bio-Paper1}
where we have shown how to construct an MLP for a full QM/MM model and how to exploit this MLP representation in a free-energy sampling approach.
Second, we combine this approach with a QM/QM refinement step for accuracy enhancement that is mediated by transfer learning toward a refined MLP representation.

In our QM/QM/MM embedding approach, selectively refining an available quantum description in a QM/MM hybrid model is achieved by defining one or more quantum cores within the quantum region. A Huzinaga-type projection-based embedding approach\cite{Hegely2016} allows us to improve the quantum description within the quantum cores. This information is then exploited in a transfer-learning approach that replaces the explicit quantum energies in the QM/QM/MM model with a machine-learning-in-MM (ML/MM) energy expression. 

Although our approach is general and will work for any QM/MM modelling approach toward physico-chemical properties, we focus on
binding free energies for protein-ligand complexes, which characterize the affinity and specificity of drug-to-target binding, making them invaluable information for drug discovery and for understanding molecular recognition in general.
We demonstrate our QM/QM/MM and transfer learning strategy for the prediction of the binding of an inhibitor (19G) to the myeloid cell leukemia 1 (MCL1) protein\cite{Friberg2012} (protein data bank entry 4HW3).

This work is structured as follows: First, we introduce our workflow, the QM/QM/MM, and the transfer machine learning approach in Section~\ref{sec:methodology}. We then discuss the effect of QM/QM/MM compared to traditional QM/MM embedding for electronic energies in Section~\ref{sec:energy-distributions}, characterize the accuracy of our MLP in Section~\ref{sec:accuracy_of_the_ml_potential}, and also discuss the effect of energy derivatives on the ML training procedure in Section~\ref{sec:energy-derivatives}. Finally, we present our QM/QM/MM estimate for the binding free energy of 19G to MCL1 in Section~\ref{sec:free_energies}.

\section{Theory of Machine Learning Assisted Multilevel Embedding\label{sec:methodology}}

\subsection{Binding Free Energies from Machine Learning Abstracted First-level Embedding}

\begin{figure}
    \centering
    \includegraphics[width=\textwidth]{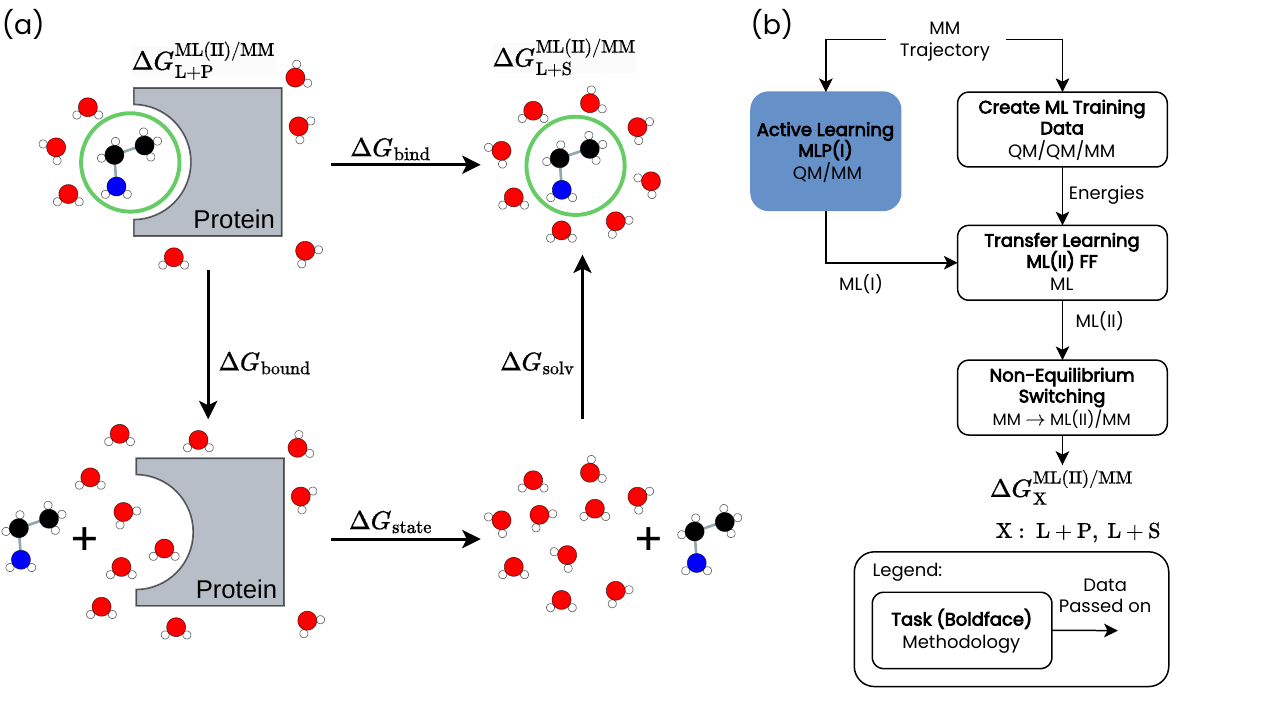}
    \caption{
    (a) Thermodynamic cycle to calculate the binding free energy of the ligand to the protein $\Delta G_\mathrm{bind}$.
    (b) Transfer learning strategy as required for the non-equilibrium switching from the MM to the ML(II)/MM force fields (FFs).}
    \label{fig:workflow}
\end{figure}

We have shown in previous work\cite{Q4Bio-Paper1} that the free energy of binding of a ligand to a protein can be calculated with an ML/MM approach by correcting the end states of classical MM-based alchemical free energy calculations\cite{Rufa2020, Tkaczyk2024}
following the thermodynamic cycle shown in Fig.~\ref{fig:workflow}(a). 
In this approach\cite{Q4Bio-Paper1}, we trained an ML potential to reproduce a QM/MM potential energy surface accurately. As a demonstration of the fidelity of the emerging potential energy hypersurface, we performed non-equilibrium switching to calculate the free energy difference of switching from the MM potential energy surface to the ML/MM potential energy surface. 
The end states for which a correction was required were the protein-ligand complex
and the solvated ligand. For details on this methodology, we refer the reader to Ref. \citenum{Q4Bio-Paper1}.

Compared to our previous work\cite{Q4Bio-Paper1}, we now improve on the MLPs by adopting more accurate quantum energies through transfer learning, as shown in Fig.~\ref{fig:workflow}(b). We here reoptimize the parameters of the MLPs from Ref.~\citenum{Q4Bio-Paper1} [from here on denoted as ML(I)] by considering highly correlated quantum chemical data obtained for quantum cores embedded in the original quantum region. 
We extracted snapshots from the MM trajectories and the structures generated during active learning used in the original workflow\cite{Q4Bio-Paper1} and calculated QM/QM/MM energies for them.
With these energies, we trained a revised MLP, denoted ML(II), and performed non-equilibrium switching from the MM potential energy surface to the ML(II)/MM potential energy surface
for the end states, that is, for the protein-ligand complex and the solvated ligand.

\subsection{Quantum Core Selection\label{qcselection}}

\begin{figure}
    \centering
    \includegraphics[width=\textwidth]{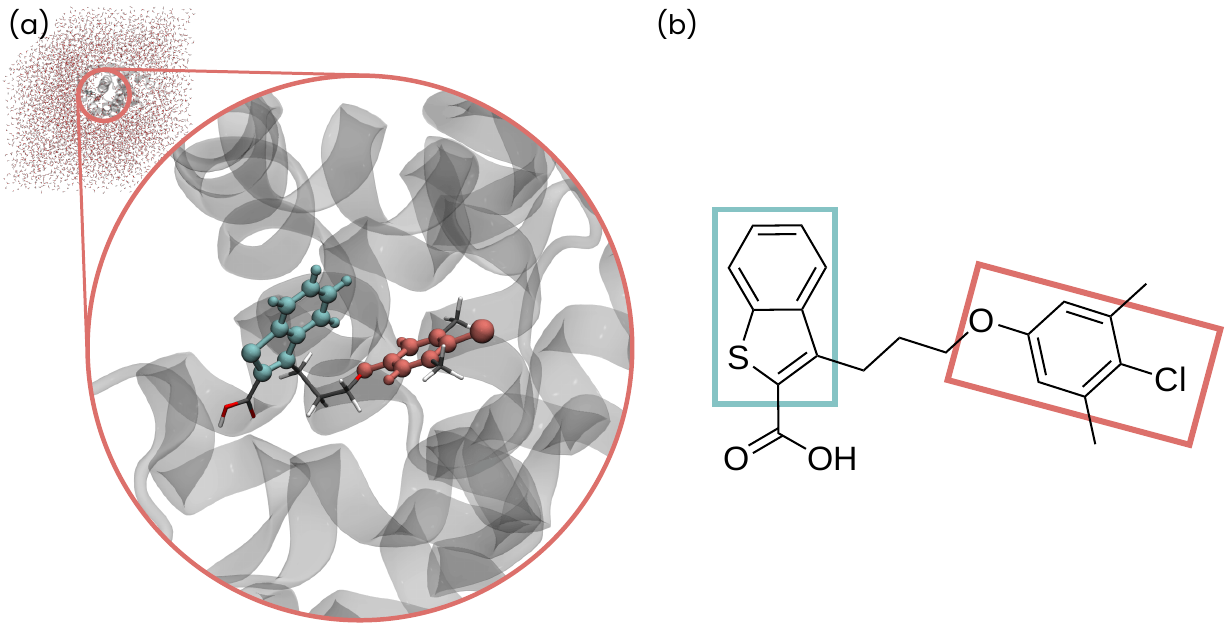}
    \caption{(a) Illustration of the protein-ligand complex. The QM region is drawn as a stick model, and the quantum cores are represented as balls and sticks. The two quantum cores are highlighted by color. (b) Lewis structure of the ligand 19G. The boxes highlight the quantum cores.}
    \label{fig:qm-cores}
\end{figure}

To demonstrate the effect of our QM/QM/MM embedding strategy on the free energy of binding, we manually selected two quantum cores from the QM region generated in Ref.~\citenum{Q4Bio-Paper1} for the MCL1-19G complex. The full system, including solvent and protein environment, is shown in Fig.~\ref{fig:qm-cores}(a), where we highlighted the QM region and color-coded the two quantum cores. Furthermore, we show the ligand's  Lewis structure with the quantum cores in Fig.~\ref{fig:qm-cores}(b). The same quantum core definition was used for all end states.

In principle, the quantum cores may be selected to include areas of the system contributing strongly to the interaction between protein and ligand since this interaction is effectively sampled during alchemical free energy simulation. However, as we will show in this paper, improving the QM region at the ligand is advantageous for increasing the overall accuracy of the QM/QM/MM calculation. Hence, since we adopt the QM region from Ref.~\citenum{Q4Bio-Paper1}, which comprises the ligand atoms, the quantum core considered in this work is also restricted to the ligand atoms. However, we note that the quantum core could, in principle, be placed anywhere within the large QM region of a QM/MM model.

The quantum cores are the separated aromatic moieties of the ligand. We selected these quantum cores because they are important for the ligand-protein interaction. 
The aromatic systems are arranged in a T-shaped structure with aromatic residues in the protein environment, leading to a favorable non-covalent interaction between ligand and protein due to the quadrupole moment of the aromatic systems. Note that the quantum cores may be chosen to be larger and include also close atoms of the protein.
With such an extended QM region, the aromatic residues of the protein could also be assigned to the quantum cores to increase the accuracy at which the interaction between the aromatic systems is described.

Furthermore, the quantum core selection can be automated by identifying parts of the electronic structure in the QM region that contribute strongly to the interaction between the ligand and the environment. For instance, an energy decomposition analysis~\cite{Kitaura1976, Bickelhaupt2003} applying computationally efficient DFT models could be used to highlight which fragments contribute strongly to the ligand-protein interaction. Alternatively, the orbitals of the QM region could be analyzed automatically\cite{Bensberg2019a, Bensberg2020} to identify the part of the electronic structure in the QM region that contributes strongly to the variation in the relative energies and, therefore, requires the most accurate description. 

We also note that it is, in principle, also possible to sample the large quantum region fully with smaller quantum cores by partitioning the system so that no atom is left in the DFT environment.

\subsection{Huzinaga-type Projection-based Embedding}

In our QM/QM/MM embedding approach, we aim to describe the quantum cores accurately by a correlated wave function method and the remaining environment with DFT. For this approach, we require a suitable embedding technique and selected Huzinaga-type embedding for this purpose.
However, other options --- such as bootstrap embedding\cite{Welborn2016, Q4Bio-Paper3b} 
--- exist and could be selected instead.

We write the total energy of our system $E_\mathrm{QM/QM/MM}$ as the sum of the energy of the molecular mechanics model $E_\mathrm{MM}$, the energy of the capped QM region $E_\mathrm{QM}$, the electrostatic interaction energy $E^\mathrm{int}_\mathrm{elec}$, and the non-electrostatic interaction energy $E^\mathrm{int}_\mathrm{LJ}$ of the QM and MM region, 
\begin{align}
    E_\mathrm{QM/QM/MM} = E_\mathrm{QM} + E_\mathrm{MM} + E^\mathrm{int}_\mathrm{elec} + E^\mathrm{int}_\mathrm{LJ}~.
\end{align}
In contrast to Ref.~\citenum{Q4Bio-Paper1}, we now increase the accuracy of $E_\mathrm{QM}$ through Huzinaga-type projection-based embedding, which allows us to apply correlated quantum chemical methods to increase the accuracy of the description of the quantum cores.

For Huzinaga-type projection-based embedding, we first calculate and localize the Kohn--Sham orbitals for the capped QM region. We then partition the full set of occupied Kohn--Sham orbitals into sets for each quantum core and one set of environment orbitals.

We write the total energy for our embedding approach as a sum of all energies for these quantum cores $E^\mathrm{QC}_I$, the Kohn--Sham energy of the environment orbitals $E^\mathrm{env}$, and the interaction $E^\mathrm{int}$ between orbital sets 
\begin{align}
    E_\mathrm{QM} = \sum_I E^\mathrm{QC}_I + E^\mathrm{env} + E^\mathrm{int}~.
\end{align}
The energy of each quantum core is given as
\begin{align}
    E^\mathrm{QC}_I = \left\langle\Psi_I\left|\hat{H}_I\right|\Psi_I\right\rangle~,
\end{align}
where $\Psi_I$ is the wavefunction of the quantum core $I$ calculated with its occupied orbitals and the full virtual orbital space of the system. The Hamiltonian operator $\hat{H}_I$ is defined in second quantization as
\begin{align}
    \hat{H}_I = \sum_{pq} h^I_{pq} a^\dagger_p a_q + \sum_{pqrs} g_{pqrs} a^\dagger_p a_q a^\dagger_r a_s~,
\end{align}
where $a^\dagger_p$ and $a_p$ denote the creation and annihilation operator for $p$, respectively. The integral $g_{pqrs}$ is the two-electron interaction integral, and $h_{pq}^I$ is the one particle integral. Only the integral $h_{pq}^I = \left\langle p \left| \hat{h}^I \right| q \right\rangle$ is modified in our embedding approach compared to a non-embedded calculation.
The operator $\hat{h}^I$ is defined as
\begin{align}
    \hat{h}^I = \hat{t} + \hat{v}_\mathrm{nuc} + \hat{v}_\mathrm{xc}^{\mathrm{nadd},I} + \hat{v}_\mathrm{J}^{I} + \hat{p}^I~,
    \label{eq:hcore}
\end{align}
where $\hat{t}$ is the kinetic energy operator, $\hat{v}_\mathrm{nuc}$ is the Coulomb potential of the QM nuclei and the MM embedding charges, and $\hat{v}_\mathrm{xc}^{\mathrm{nadd}, I}$ is the non-additive exchange--correlation potential
\begin{align}
    \hat{v}_\mathrm{xc}^{\mathrm{nadd},I}(\pmb{r}) = \left.\frac{\delta E_\mathrm{xc}[\rho(\pmb{r})]}{\delta \rho(\pmb{r})}\right|_{\rho=\rho_\mathrm{tot}} - \left.\frac{\delta E_\mathrm{xc}[\rho(\pmb{r})]}{\delta \rho(\pmb{r})}\right|_{\rho=\rho_I}~.
\end{align}
$E_\mathrm{xc}[\rho]$ denotes the exchange--correlation functional evaluated with the electron density~$\rho$, $\rho_\mathrm{tot}$ denotes the total density of the QM region, and $\rho_I$ denotes the density of the quantum core $I$.
The Coulomb interaction with the electrons not in quantum core $I$ is given by $\hat{v}_\mathrm{J}^I$
\begin{align}
    \hat{v}_\mathrm{J}^I = \sum_{K \neq I} \int \frac{\rho_K(\pmb{r}^\prime)}{\left|\pmb{r} - \pmb{r}^\prime\right|} \mathrm{d}\pmb{r}^\prime~.
\end{align}
The operator $\hat{p}^I$ is a projection operator ensuring orthogonality between orbital sets. It is defined as 
\begin{align}
    \hat{p}^I = -\sum_{J\neq I} \hat{P}_J \left(\hat{f}^I - \epsilon_\mathrm{shift} \right) - \left(\hat{f}^I - \epsilon_\mathrm{shift} \right) \sum_{J\neq I} \hat{P}_J~,
\end{align}
where $\hat{P}_J = \sum_{i \in \mathrm{occ}} \left| i\right\rangle\left\langle i\right|$ is the projection operator on the occupied orbitals of the environment fragment (quantum core or DFT environment) $J$, $\hat{f}^I$ is the embedded Fock operator of the quantum core $I$, and $\epsilon_\mathrm{shift}$ is a constant positive shift~\cite{Chulhai2018, Bensberg2020a}.
Note that this shift prevents a shift of environment orbitals to lower energies (producing a violation of the Aufbau principle) but does not affect the physics of the embedding procedure. 

The interaction energy $E^\mathrm{int}$ can be written as
\begin{align}
\begin{split}
    E^\mathrm{int} = -\sum_I \left\langle\Psi_I\left| \hat{v}_\mathrm{xc}^{\mathrm{nadd},I} \right| \Psi_I\right\rangle - \sum_{I < J} \int\int \frac{\rho_I(\pmb{r}^\prime)\rho_J(\pmb{r})}{\left|\pmb{r}^\prime - \pmb{r}\right|}\mathrm{d}\pmb{r}^\prime\mathrm{d}\pmb{r} + E_\mathrm{xc}^\mathrm{nadd}~,
\end{split}
\end{align}
where we subtracted the contribution of $\hat{v}_\mathrm{xc}^{\mathrm{nadd},I}$ from Eq.~(\ref{eq:hcore}) because it is already included in the non-additive exchange--correlation energy $E_\mathrm{xc}^\mathrm{nadd}$
\begin{align}
    E_\mathrm{xc}^\mathrm{nadd} = E_\mathrm{xc}[\rho_\mathrm{tot}(\pmb{r})] - \sum_I E_\mathrm{xc}[\rho_I(\pmb{r})] - E_\mathrm{xc}[\rho_\mathrm{env}(\pmb{r})]~.
\end{align}
Furthermore, the Coulomb interaction between the quantum cores is already effectively included in the operator $\hat{v}_\mathrm{J}^I$ twice. Therefore, we subtracted the Coulomb interaction between quantum cores again to avoid double counting. Note that in practice, only the densities of the Kohn--Sham DFT orbitals selected for the fragments are used to evaluate the interaction term $E^\mathrm{int}$ since densities or density matrices for the correlated wavefunction of the quantum core may not always be available.

Moreover, we avoid the embedded self-consistent field procedure common in projection-based embedding to obtain the Hartree--Fock orbitals for the quantum cores but rely directly on the initial Kohn--Sham orbitals. This not only simplifies our embedding approach significantly from a technical perspective but also allows us to use DFT orbitals for the correlated wave function region, which are often considered superior to Hartree--Fock orbitals~\cite{Fang2016, Fang2017, Bertels2021}.

\subsection{Transfer Machine Learning of Second-level Embedding Energies\label{sec:transferlearning}}

The second key component of our approach is a transfer learning step which connects the accurate quantum-core energies with the global potential energy surface represented in an ML/MM model.

In a brute-force approach (that is, without transfer learning), the structure-energy relation of the reference method would be directly learned by an ML potential \cite{Behler2007, Behler2016, Bartok2017, Noe2020, Kaeser2023}. The ML potential training is based on representative structures and their total energies and gradients. However, calculating energies and gradients with multi-level embedding approaches requires high computational effort. In some cases, an implementation for the gradients may be unavailable. Therefore, training the QM/QM/MM energies $E_\mathrm{QM/QM/MM}$ from scratch can be challenging. Hence, we start here from a pre-trained (system-focused) ML potential, such as the one obtained in our previous work \cite{Q4Bio-Paper1}. Typically, many more QM/MM than QM/QM/MM data points can be generated. These data points have already been generated to learn the QM/MM potential energy surface $E_\mathrm{QM/MM}$ in our previous work.

A refinement step will now be necessary to incorporate higher accuracy quantum data. Only the energy difference between QM/QM/MM and QM/MM $\Delta E_\mathrm{QM/QM/MM-QM/MM}$ needs to be learned from the QM/QM/MM data,
\begin{align}
E_\mathrm{QM/QM/MM} = E_\mathrm{QM/MM} + \Delta E_\mathrm{QM/QM/MM-QM/MM}\ .
\end{align}
This correction is typically smoother than the full QM/QM/MM potential energy surface. Hence, fewer data points are required for sufficient sampling. The correction can be learned by a separate ML potential in a so-called $\Updelta$-learning approach \cite{Ramakrishnan2015, Zaspel2019, Dral2020}. However, the inference requires then the evaluation of two MLPs. Training such a second MLP from scratch will again be difficult (for the same reason as highlighted before for the base MLP: the number of QM/QM/MM data points is small and/or no gradients are available for QM/QM/MM). Therefore, we apply transfer learning \cite{Smith2019, Kaeser2022, Chen2023, Zaverkin2023, Dral2023}, where an MLP is first trained on QM/MM energies and gradients. Subsequently, some weight parameters of this MLP are fine-tuned to transfer its knowledge to the prediction of QM/QM/MM energies.

\subsection{Practical Realization of Multi-level Learning\label{sec:technicalitiesoflearning}}

In the MLP approach toward a QM/MM potential energy surface described in Ref. \citenum{Q4Bio-Paper1}, the MM energies $E_\mathrm{MM}$ and $E_\mathrm{MM}^\mathrm{int}$ are not learned since they can be calculated efficiently with the MM force field. We also separated in that approach electrostatic interactions between mixed QM--MM atom pairs from the energy to be learned,
\begin{align}
\begin{split}
&E_\mathrm{QM/MM} = E_\mathrm{QM/MM}^\mathrm{ML} + E_\mathrm{MM} + E_\mathrm{MM}^\mathrm{int}\\
&+ \left(\sum_{I\in Q}\sum_{A\in E}\frac{q_I q_A}{|\pmb{R}_I - \pmb{R}_A|}\right) + \left(\sum_{m=1}^{N_\mathrm{elem}}\sum_{n=1}^{N_\mathrm{atom}^{m}}E_\mathrm{QM}^{\mathrm{atomic},m}\right)\ .
\end{split}
\end{align}
The atomic charges $q_I$ at positions $\pmb{R}_I$ of the QM atoms $I$ and the charges $q_A$ at positions $\pmb{R}_A$ of MM atoms $A$ are taken from the MM force field in this work. Moreover, the atomic contributions to the QM energy $E_\mathrm{QM}^{\mathrm{atomic},m}$ are separated to simplify training. Their values depend only on the chemical element type $m$. In this way, the ML potential energy $E_\mathrm{QM/MM}^\mathrm{ML}$ represents only the QM energy (without atomic contributions) and the difference in the electrostatic QM--MM interaction between QM/MM and pure MM representations.

In transfer learning, the MLP pre-trained on the QM/MM data is fine-tuned on the energy $E_\mathrm{QM/MM}^\mathrm{ML} + \Delta E_\mathrm{QM/QM/MM-QM/MM}^\mathrm{ML}$. Since the QM/MM and QM/QM/MM potential energy surfaces  can be shifted, the mean shift from QM/MM energies (without atomic contributions) to QM/QM/MM energies in the training data is handled separately,
\begin{align}
\begin{split}
&E_\mathrm{QM/QM/MM} = E_\mathrm{QM/MM}^\mathrm{ML} + \Delta E_\mathrm{QM/QM/MM-QM/MM}^\mathrm{ML} + E_\mathrm{MM} + E_\mathrm{MM}^\mathrm{int}\\
&+ \left(\sum_{I\in Q}\sum_{A\in E}\frac{q_I q_A}{|\pmb{R}_I - \pmb{R}_A|}\right) + \left(\sum_{m=1}^{N_\mathrm{elem}}\sum_{n=1}^{N_\mathrm{atom}^{m}}E_\mathrm{QM}^{\mathrm{atomic},m} + \overline{\Delta E}_\mathrm{QM/QM/MM-QM/MM}^m\right)\ .
\end{split}
\end{align}
The shifts $\overline{\Delta E}_\mathrm{QM/QM/MM-QM/MM}^m$ can be obtained in an element-dependent form by a least squares fit of the differences in the QM/MM and QM/QM/MM training energies with respect to the stoichiometries. 

We note that the specific choice for the shifts is not crucial for the whole training process. It eases the transfer learning, but any deficiencies of the shift (even a complete lack of the shift) can be compensated by the training process in the long run.

In conclusion, $\Delta E_\mathrm{QM/QM/MM-QM/MM}^\mathrm{ML}$ represents only relative energy differences between QM/MM and QM/QM/MM approaches, centered around zero. Consequently, the MLP modifications during the training of the QM/QM/MM data should be as small as possible, enabling a sufficient representation by rather small amounts of data. For a leaner notation in the following, we introduce the shorthand notation $E_\mathrm{ML}=E_\mathrm{QM/MM}^\mathrm{ML} + \Delta E_\mathrm{QM/QM/MM-QM/MM}^\mathrm{ML}$.

Optimizing the weight parameters only on the QM/QM/MM corrections can still lead to a loss of previous QM/MM information about the general shape of the potential energy surface. To mitigate this loss, a fraction of the MLP weight parameters are not modified. For the other weights, the state of the CoRe optimizer \cite{Eckhoff2023, Eckhoff2024} at the last training step of the QM/MM MLP is reloaded and continued, which includes learning rates individually adapted for each weight.

\section{Computational Details}

\subsection{Multilevel Embedding}
For the QM/QM/MM embedding, the initial DFT calculation for the full QM region applied the Perdew, Burke, and Ernzerhof exchange--correlation functional PBE\cite{Perdew96}, the D3 dispersion correction\cite{Grimme2010a} with Becke--Johnson damping (BJ)\cite{Grimme2011}, and the def2-TZVP basis set\cite{Ahlrich2005}. The occupied core and valence orbitals were then localized separately, following the intrinsic bond orbital approach\cite{Knizia2013}, and partitioned into subsystems. Occupied orbitals were assigned to the environment if their Mulliken population exceeded $0.4$ on the environment atoms. Otherwise, they were assigned to the quantum core on which they had the largest population. The electronic structure of the quantum cores was described by domain-based local pair natural orbital coupled cluster with singles, doubles, and semi-canonical perturbative triples excitations [DLPNO-CCSD(T$_0$)]\cite{Riplinger2013a, Riplinger2016} using ``NormalPNO'' thresholds\cite{Liakos2015}. The shift $\epsilon_\mathrm{shift}$ was chosen to be $1.0~\si{au}$. All QM calculations were carried out with the Serenity program\cite{Serenity2018, Niemeyer2022, Artiukhin2025}.

The QM calculations were embedded in an MM environment by electrostatic embedding, the solvent, and the counter ions. The parameters for the MM environment were taken from the Amber99SB-ILDN\cite{LindorffLarsen2010} force field and TIP3P water model\cite{Jorgensen1983}. The  General Amber Force Field (version 2.11)\cite{Wang2004, Wang2005} provided the Lennard--Jones parameters for the non-covalent interaction between ligand atoms and the MM environment. The program Swoose\cite{Brunken2020, zenodo-archive} calculated all force field terms.

The quantum cores were chosen according to the discussion in section \ref{qcselection} and are illustrated in Fig.~\ref{fig:qm-cores}. The first quantum core [blue in Fig.~\ref{fig:qm-cores}(a)] consisted of the benzothiophene moiety of the ligand. The second quantum core [red in Fig.~\ref{fig:qm-cores}(b)] consisted of the ligand's mesityl chloride moiety (excluding the methyl groups).
We calculated the QM/QM/MM energies for all structures of MCL1-19G and 19G, which showed a QM/MM energy within $150~\si{kJ.mol^{-1}}$ of the median of the respective QM/MM energy distribution.

We will refer to energies calculated with our QM/QM/MM approach as $E_\mathrm{QM/QM/MM}$, to energies calculated with the QM/MM (PBE-D3(BJ)/def2-SVP-in-Amber) approach in Ref.~\citenum{Q4Bio-Paper1} as $E_\mathrm{QM/MM}$, and to energies obtained from the initial DFT calculation (PBE-D3(BJ)/def2-TZVP-in-Amber) for the full QM region during the embedding procedure as $E_\mathrm{QM2/MM}$.

\subsection{Transfer Learning}

The MLP was based on an ensemble of high-dimensional neural network potentials (HDNNPs) \cite{Behler2007, Behler2021}. HDNNPs employ element-embracing atom-centered symmetry functions (eeACSFs) as their input features \cite{Eckhoff2023}. As the QM/QM/MM MLP was trained based on the QM/MM MLP of Ref.~\citenum{Q4Bio-Paper1}, the eeACSF parameters and neural network architectures were chosen as in that reference. Furthermore, we continued to apply an ensemble size of 10 and an uncertainty scaling factor of $c=2$ \cite{Eckhoff2023}.

The set of QM/QM/MM reference energies was split randomly into $90\%$ training and $10\%$ test data for each ensemble member individually unless stated otherwise. Transfer learning employed the CoRe optimizer \cite{Eckhoff2023, Eckhoff2024} (version 1.1.0) \cite{Eckhoff2024a} to minimize the difference in energy predictions and respective training data. The hyperparameters of the CoRe optimizer were taken from Ref.~\citenum{Q4Bio-Paper1}. The optimizer state of the last training step in learning the QM/MM potential energy surface was loaded, and the training was continued. In each of the 300 training steps, all training energies were exploited. The weight parameters for the standardization of the input features \cite{Eckhoff2023} and those of the first hidden layer of the neural networks were not varied in transfer learning.

Transfer learning and inference of the MLPs was carried out with the lMLP software (version 2.0.0) \cite{Eckhoff2024b}. The program lMLP is available on GitHub, PyPI, and included in the Zenodo archive\cite{zenodo-archive}.

\subsection{End-state correction simulations}
The non-equilibrium (NEQ) switching simulations for the end-state corrections in the thermodynamic cycle were run as described in Ref.~\citenum{Q4Bio-Paper1}. In this protocol, 150 structures were randomly selected from the MM end states. For these structures, the force field was switched slowly from MM to ML(II)/MM over $10~\si{ps}$, using an NPT ensemble and the same conditions employed in the MM simulations. After the forward switch, we resampled the structures to represent best the equilibrium distribution of the ML(II)/MM potential energy surface, propagated the structures for another $10~\si{ps}$, and then switched the potential energy surface again from ML(II)/MM to MM. This NEQ switching procedure was repeated six times for the protein-ligand complex and the solvated ligand to estimate the uncertainty from the sampling during NEQ switching.

We estimate the uncertainty $\delta G^\mathrm{ML(II)/MM}_\mathrm{bind}$ for the binding free energy with the error from the underlying MM calculation and the sampling error from the NEQ simulations and Multistate Bennett acceptance ratio (MBAR)\cite{Shirts2003} as
\begin{align}
    \delta G^\mathrm{ML(II)/MM}_\mathrm{bind} = \sqrt{\left(\delta G^{\mathrm{MBAR}}_\mathrm{L+P}\right)^2 + \left(\delta G^{\mathrm{MBAR}}_\mathrm{L+S}\right)^2 + \left(2\sigma^\mathrm{ML(II)/MM}_\mathrm{bind}\right)^2}~.
\end{align}
Here, $\delta G^{\mathrm{MBAR}}_\mathrm{L+S}$ and $\delta G^{\mathrm{MBAR}}_\mathrm{L+P}$ denote the MBAR error estimates for the solvated ligand and the protein-ligand complex, respectively, and $\sigma^\mathrm{ML(II)/MM}_\mathrm{bind}$ denotes the standard deviation of the 36 binding free energy estimations (6 $\times$ 6 for each end state).

\section{Results for Binding Free Energies from Hierarchical Quantum Hybrid Models}

\subsection{Energy Distributions in Quantum Hybrid Models\label{sec:energy-distributions}}

Since we aim at non-equilibrium switching simulations to approximate the free energy difference between the QM/MM and the QM/QM/MM potential energy surfaces, we initially investigated the difference in relative energies between both approaches. The initial QM/MM model and our QM/QM/MM approach differ in two aspects. (i) We employed the def2-TZVP basis set in the QM/QM/MM calculations, but only the smaller def2-SVP basis set in the initial QM/MM calculation. (ii) The QM/QM/MM approach described two quantum cores with DLPNO-CCSD(T$_0$).

\begin{figure}[tb!]
    \centering
    \includegraphics[width=\textwidth]{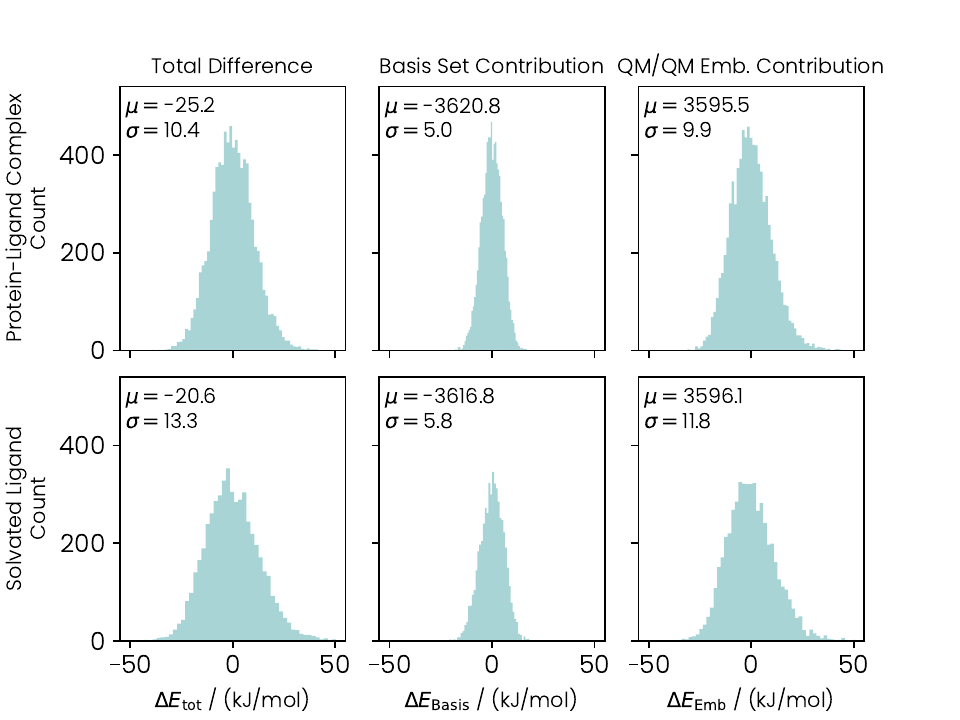}
    \caption{Distributions of the energy differences of $\Delta E_\mathrm{tot}$, $\Delta E_\mathrm{Basis}$, and $\Delta E_\mathrm{Emb}$ for the protein-ligand complex and the ligand in solution. The distributions are shifted by their mean for clarity. The standard deviations ($\sigma$) and means ($\mu$) are given in $\si{kJ.mol^{-1}}$.}
    \label{fig:energy-differences}
\end{figure}

To disentangle the effect of the basis set and the QM/QM embedding, we plotted the total energy difference $\Delta E_\mathrm{tot} = E_\mathrm{QM/QM/MM} - E_\mathrm{QM/MM}$, the difference in the energies caused by the basis set $\Delta E_\mathrm{Basis} = E_\mathrm{QM2/MM} - E_\mathrm{QM/MM}$, and the difference in the energies caused by the QM/QM embedding $\Delta E_\mathrm{Emb} = E_\mathrm{QM/QM/MM} - E_\mathrm{QM2/MM}$ in  Fig.~\ref{fig:energy-differences}.

The distributions for the energy differences for both end states (protein-ligand complex and solvated ligand) are symmetric and almost of Gaussian shape. Overall, the distribution for the total energy differences $\Delta E_\mathrm{tot}$ are relatively narrow with standard deviations of $10.4~\si{kJ.mol^{-1}}$ and $13.3~\si{kJ.mol^{-1}}$ for the protein-ligand complex and the solvated ligand, respectively. The standard deviations for $\Delta E_\mathrm{Emb}$ are $9.9~\si{kJ.mol^{-1}}$ and $11.8~\si{kJ.mol^{-1}}$ for the respective end states, which accounts for most of the variance in $\Delta E_\mathrm{tot}$. By contrast, increasing the basis set size from def2-SVP to def2-TZVP results only in a variance of $5.0~\si{kJ.mol^{-1}}$ and $5.8~\si{kJ.mol^{-1}}$ for the protein-ligand complex and the solvated ligand, respectively. Since we need to calculate the energy difference between both end states, the mean differences of the energy distributions provide a first hint for the effect on the final free energy differences. Here, the means of $\Delta E_\mathrm{tot}$ for the protein-ligand complex and solvated ligand differ by $-4.6~\si{kJ.mol^{-1}}$, suggesting that the QM/QM embedding stabilizes the bound state slightly. The majority of this mean shift ($-4.0~\si{kJ.mol^{-1}}$) is caused by the improved basis set, while the remaining $-0.6~\si{kJ.mol^{-1}}$ can be traced back to the QM/QM embedding.

\subsection{Accuracy of the Machine Learning Potential\label{sec:accuracy_of_the_ml_potential}}

For both systems, the protein-ligand complex and the solvated ligand, a separate MLP is trained to achieve the best accuracy-cost ratio for inference. Transfer learning is based on $8417$ reference conformers for the protein-ligand complex and $5285$ for the solvated ligand, while each reference conformer contains $N_Q=44$ QM atoms. The ranges of the target reference QM/QM/MM energies per QM atom $E_\mathrm{ML}^\mathrm{ref}\,N_Q^-1$ are $8.649$ and $8.756,\mathrm{kJ\,mol}^{-1}$, respectively, while the standard deviations are $0.845$ and $0.857\,\mathrm{kJ\,mol}^{-1}$.

The resulting energy root mean square errors (RMSEs) are one order of magnitude smaller than the corresponding energy ranges and significantly smaller than the standard deviations, confirming high fidelity
(Table \ref{tab:MLP_accuracy}). Training and test data show similar errors indicating the absence of overfitting. Ensembling of ten HDNNPs in an MLP reduces the RMSE values compared to the individual predictions, especially for the solvated ligand.

The QM/QM/MM MLP RMSEs are $5\%$ and $14\%$ lower than those of the base QM/MM MLPs of the protein-ligand complex and for the solvated ligand, respectively. Whereas the standard deviations of the training data for the base potential was $3.6~\si{kJ.mol^{-1}}$ for the protein-ligand complex and $5.9~\si{kJ.mol^{-1}}$ for the solvated ligand, the standard deviations for the training data used here turned out to be $0.8~\si{kJ.mol^{-1}}$ and $0.9~\si{kJ.mol^{-1}}$, respectively. Therefore, the increased accuracy can be attributed to the narrower energy distribution of the training data, a consequence of the restriction to structures within $150~\si{kJ.mol^{-1}}$ of the median energy.

\begin{table*}[htb!]
\caption{
RMSEs of target QM/QM/MM energies $E_\mathrm{ML}$ for HDNNPs prior to ensembling and of HDNNP ensembles. The mean and standard deviations of ten individual HDNNPs are provided for training and test data. The HDNNP ensemble was applied to all data.}
\begin{center}
\begin{tabular}{lrr}
\hline
RMSEs before ensembling & MCL1-19G & 19G\\
\hline
$E_\mathrm{ML}^\mathrm{train}N_Q^{-1}\,/\,\mathrm{kJ\,mol}^{-1}$ & $0.159\pm0.004$ & $0.218\pm0.005$\\
$E_\mathrm{ML}^\mathrm{test}N_Q^{-1}\,/\,\mathrm{kJ\,mol}^{-1}$ & $0.169\pm0.007$ & $0.243\pm0.009$\\
\hline
Ensemble RMSEs\\
\hline
$\overline{E}_\mathrm{ML}N_Q^{-1}\,/\,\mathrm{kJ\,mol}^{-1}$ & $0.150$ & $0.205$\\
\hline
\end{tabular}
\end{center}
\label{tab:MLP_accuracy}
\end{table*}

\begin{figure*}[htb!]
\centering
\includegraphics[width=0.5\textwidth]{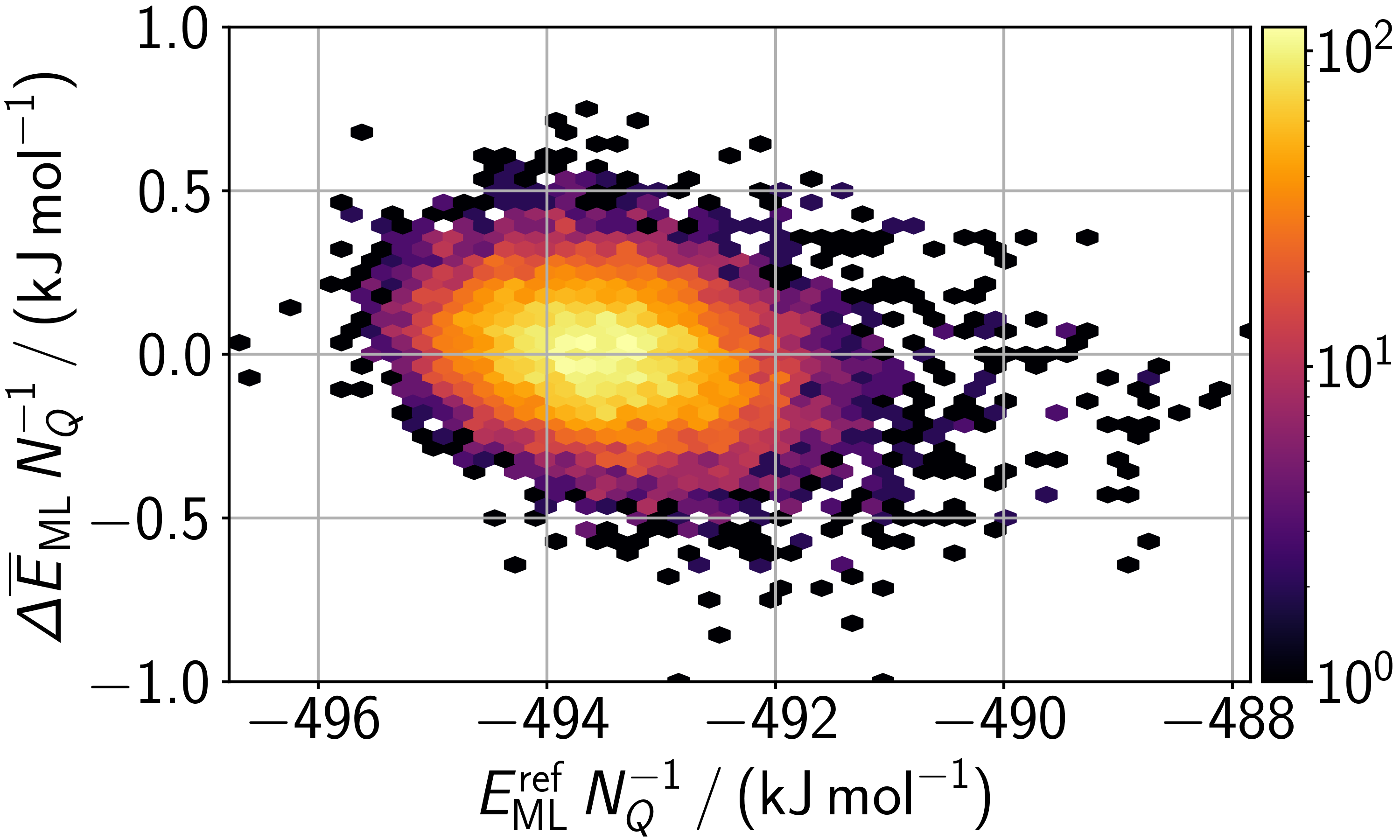}
\caption{Difference in the ensemble prediction of target QM/QM/MM energies $\Delta\overline{E}_\mathrm{ML}$ from the reference data $E_\mathrm{ML}^\mathrm{ref}$ as a function of the reference data $E_\mathrm{ML}^\mathrm{ref}$ for both systems, MCL1-19G and 19G, normalized by the number of QM atoms $N_Q$. Color in this hexagonal binning plot visualizes the number of data points in a hexagon. Three outlier data points are outside the shown error range.}\label{fig:MLP_error_distribution}
\end{figure*}

\begin{figure*}[htb!]
\centering
\includegraphics[width=0.5\textwidth]{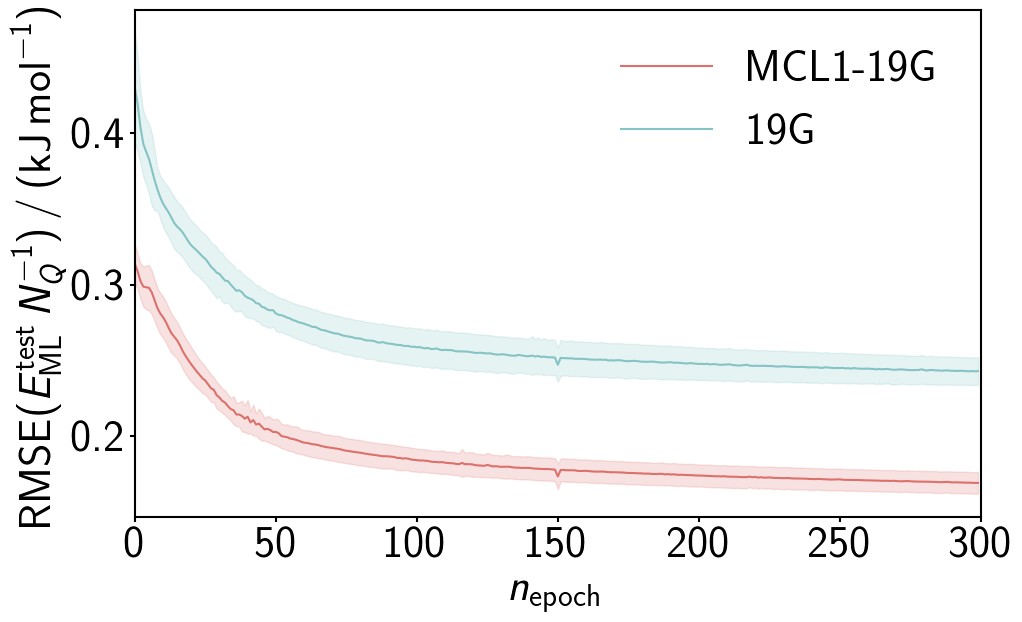}
\caption{Transfer learning progress of $\mathrm{RMSE}(E_\mathrm{ML}^\mathrm{test}\,N_Q^{-1})$ or both systems, MCL1-19G and 19G, as a function of the training epoch $n_\mathrm{epoch}$. The solid line represents the mean of ten individual HDNNPs and the shaded area shows the standard deviation.}\label{fig:MLP_training}
\end{figure*}

The error distribution in Figure \ref{fig:MLP_error_distribution} shows that the errors are well centered around zero and that the majority of data points show absolute errors smaller than $0.5\,\mathrm{kJ\,mol}^{-1}$ per QM atom. The training process in Figure \ref{fig:MLP_training} highlights the smooth convergence of the transfer learning approach as well as of the CoRe optimizer. Moreover, it reveals that the relative difference of the QM/QM/MM and QM/MM potential energy surfaces is larger for the solvated ligand than for the protein-ligand complex, since the initial energy RMSE is larger for the solvated ligand.

\subsection{Effect of Energy Gradients and Amount of Training Data on Transfer Learning\label{sec:energy-derivatives}}

Atomic forces, i.e., negative energy gradients, are not available in our current implementation of the QM/QM/MM energies. However, training on forces may significantly improve the learning process because they provide information about the local shape of the potential energy surface. To probe the extent to which forces can affect the resulting accuracy of a transfer learned MLP, we constructed an embedding approach where the QM region is described by an MM force field. In this way, we can train a base model using $10^4$ snapshots from the MM trajectory on MM/MM energies and apply transfer learning to the QM/MM potential energy surface, while in both cases, energy gradients are available. In addition, we can analyze the quality of the resulting force predictions for transfer learning, which utilizes only energies. Furthermore, direct learning of the QM/MM data also allows us to obtain a reference for the resulting accuracy.

Transfer learning only on energies yields similar RMSEs for the energies $E^\mathrm{QM/MM}$, by contrast to transfer learning on energies and forces (Table \ref{tab:TL_gradients}). However, the resulting RMSE for the atomic force components of the QM atoms $F_{\alpha,n(Q)}^\mathrm{QM/MM}$ is about twice as high for training only on energies compared to training on energies and forces. A higher force RMSE for the former learning case is expected as much less information about the potential energy surface is available. Each of the $9\cdot10^3$ training conformers provides only one energy value, while $3\cdot44$ force components of QM atoms are available. For the RMSE of the atomic force component of the MM atoms represented by the MLP $F_{\alpha,n(E^\prime)}^\mathrm{QM/MM}$, the difference between the learning cases is not as large. However, the absolute values of these force errors are, in general, one order of magnitude smaller than those of $F_{\alpha,n(Q)}^\mathrm{QM/MM}$, which leads to less emphasis on $F_{\alpha,n(E^\prime)}^\mathrm{QM/MM}$ during training.

\begin{table*}[htb!]
\caption{RMSEs of QM/MM energies $E^\mathrm{QM/MM}$ and QM/MM atomic force components of QM atoms $F_{\alpha,n(Q)}^\mathrm{QM/MM}$ and MM atoms represented by the MLP $F_{\alpha,n(E^\prime)}^\mathrm{QM/MM}$ for HDNNP ensembles evaluated on all data, i.e., training and test data. The HDNNP ensembles were obtained by training on pure MM data and transfer learning to QM/MM data, while either only energies or energies and forces were exploited in transfer learning. Direct training on QM/MM data is given as a reference. The underlying $10^4$ reference conformers were obtained from MM sampling during AFE simulations.}
\begin{center}
\begin{tabular}{llrrr}
\hline
Learning & System & \multicolumn{1}{l}{RMSE} & \multicolumn{1}{l}{RMSE} & \multicolumn{1}{l}{RMSE}\\
& & \multicolumn{1}{l}{$\overline{E}^\mathrm{QM/MM}N_Q^{-1}$} & \multicolumn{1}{l}{$\overline{F}_{\alpha,n(Q)}^\mathrm{QM/MM}$} & \multicolumn{1}{l}{$\overline{F}_{\alpha,n(E^\prime)}^\mathrm{QM/MM}$}\\
& & \multicolumn{1}{l}{$/\,\mathrm{kJ\,mol}^{-1}$} & \multicolumn{1}{l}{$/\,\mathrm{kJ\,mol}^{-1}\,\text{\AA}^{-1}$} & \multicolumn{1}{l}{$/\,\mathrm{kJ\,mol}^{-1}\,\text{\AA}^{-1}$}\\
\hline
Transfer & MCL1-19G & 0.150 & 22.1 & 1.01\\
$E$ & 19G & 0.195 & 23.7 & 1.81\\
\hline
Transfer & MCL1-19G & 0.145 & 11.3 & 0.94\\
$E$ and $F$ & 19G & 0.212 & 12.8 & 1.73\\
\hline
Direct & MCL1-19G & 0.117 & 9.0 & 0.81\\
& 19G & 0.161 & 10.1 & 1.48\\
\hline
\end{tabular}
\end{center}
\label{tab:TL_gradients}
\end{table*}

Comparing the results of transfer learning to direct learning shows that the energy RMSE increases by around $26\%$ (Table \ref{tab:TL_gradients}). Here, the same number of reference conformers are employed, but still the flexibility of the ML potential parameters and the number of updates during transfer learning are significantly lower than for direct learning. For $F_{\alpha,n(Q)}^\mathrm{QM/MM}$, transfer learning energies and forces also yields an increase of the RMSE of about $26\%$ compared to direct learning. The increase is about $16\%$ for $F_{\alpha,n(E^\prime)}^\mathrm{QM/MM}$. Still, the RMSEs are decent for the following Alchemical Free Energy (AFE) simulations, even if the transfer learning is only based on energies since the RMSEs are in a similar range as those of previously employed HDNNPs in simulations on pure QM potential energy surfaces\cite{Eckhoff2019, Eckhoff2021a}.

\begin{figure*}
    \centering
    \includegraphics[width=0.7\textwidth]{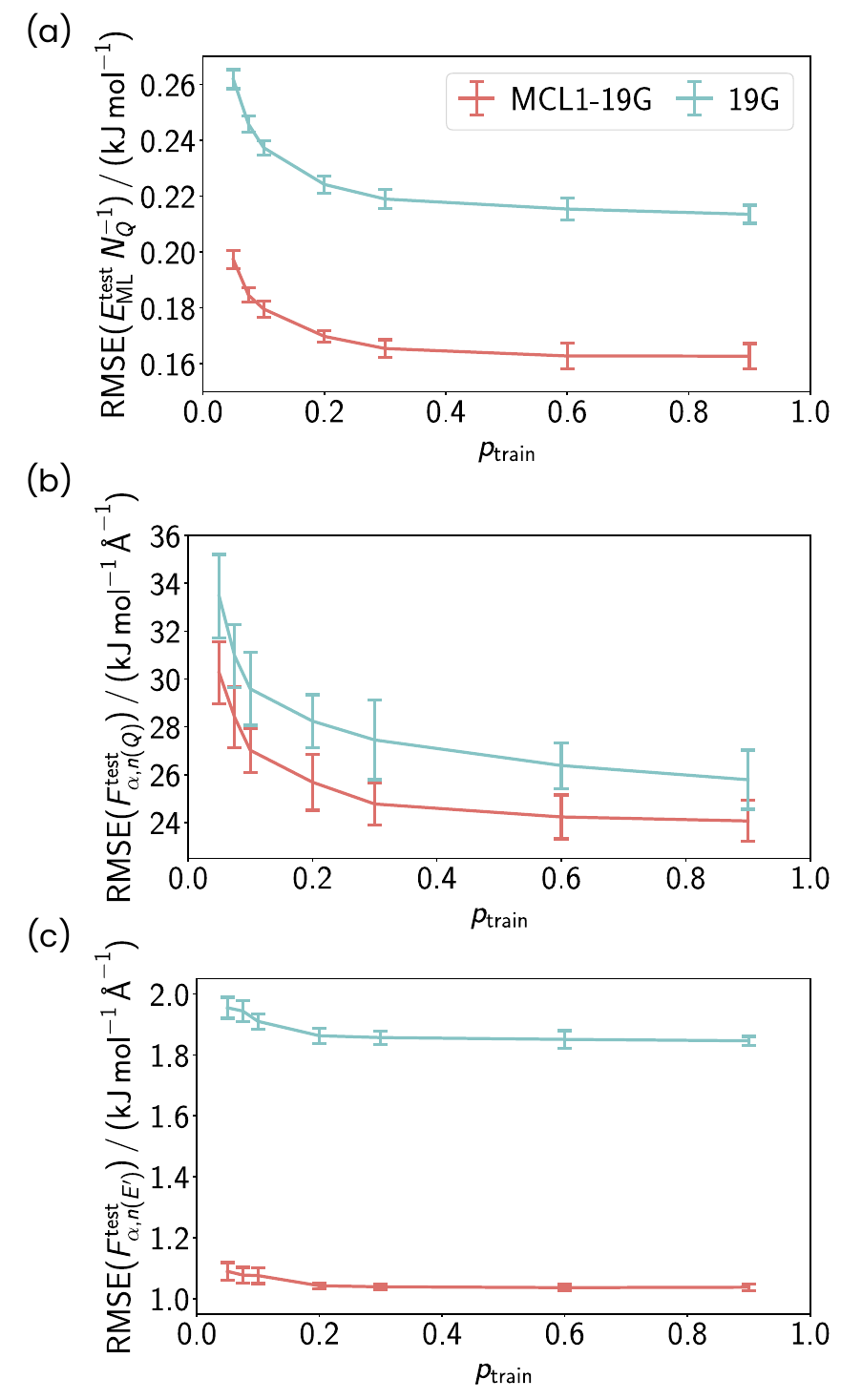}
    \caption{Test set RMSEs of (a) QM/MM energies $E_\mathrm{ML}^\mathrm{test}$ and (b) QM/MM atomic force components of QM atoms $F_{\alpha,n(Q)}^\mathrm{test}$ and (c) MM atoms represented by the ML potential $F_{\alpha,n(E^\prime)}^\mathrm{test}$ as a function of the training fraction $p_\mathrm{train}$. The lines represent the mean of ten individual HDNNPs predictions, i.e., ensembling is not applied, and the error bars show the standard deviations. The HDNNPs were obtained by training on pure MM data and transfer learning to different amounts of QM/MM data, while only energies were utilized in transfer learning. The underlying $10^4$ reference conformers were obtained from MM sampling during AFE simulations.}\label{fig:TL_training_fraction}
\end{figure*}

To understand how many energy values for training are required in transfer learning for the representation of the conformation space sampled by MM in AFE simulations, the training data fraction $p_\mathrm{train}$ was varied. We observe that reducing the training fraction of the $10^4$ reference conformers from $0.9$ to $0.3$ increases the resulting test RMSEs of energies and forces only little (Figures \ref{fig:TL_training_fraction} (a) to (c)). Decreasing the number of training conformers to $10^3$ leads to increases in the $E^\mathrm{QM/MM}$ and $F_{\alpha,n(Q)}^\mathrm{QM/MM}$ RMSEs of about $12\%$ compared to utilizing $9\cdot10^3$ conformers. If the number is further decreased to $0.5\cdot10^3$, the RMSEs will increase by about $25\%$. Consequently, the MLP accuracy is still fine, even if only about a tenth of the conformers and only energies are employed. The smoothness of the transfer learned potential energy surface is probed in the AFE simulations presented in the next section. These simulations can finally confirm that the forces of the QM atoms, which are predicted by an MLP obtained in transfer learning only on energies, lead to reasonable dynamics. 

\subsection{Free Energy Corrections from Non-Equilibrium Switching\label{sec:free_energies}}

\begin{figure}[t]
    \centering
    \includegraphics[width=0.7\textwidth]{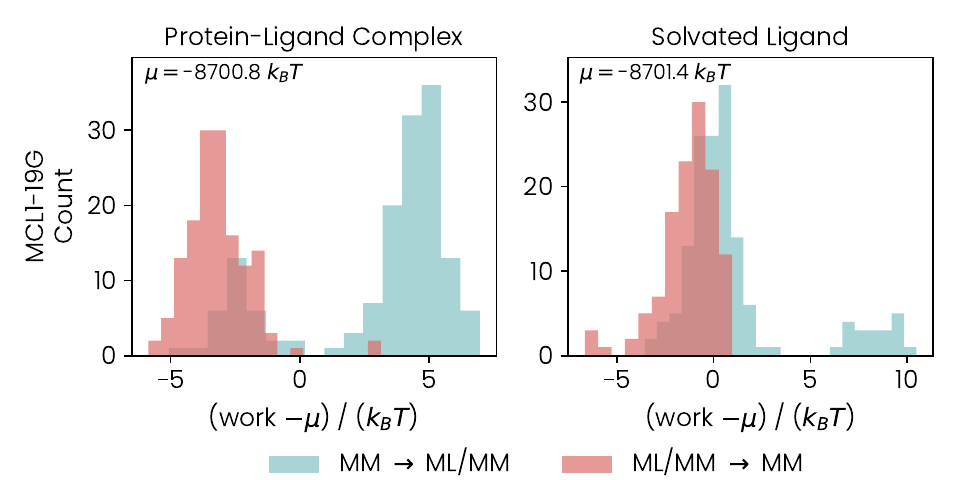}
    \caption{Work distributions from NEQ switching simulations. The results for the protein-ligand complex are shown in the left plot and those for the solvated ligand in the right plot.}
    \label{fig:work-distributions}
\end{figure}

The work distributions for the first out of six NEQ switching simulations are shown in Fig.~\ref{fig:work-distributions} for the protein-ligand complex and the solvated ligand.
For both end states, we observe two distinct peaks in the work distribution. In Ref.~\citenum{Q4Bio-Paper1}, these peaks have already been related to the preference of the carboxylic acid to form hydrogen bonds to sulfur or the solvent. On the MM potential energy surface, the ligand forms a hydrogen bond with a neighboring sulfur atom for some conformations. These hydrogen bonds are fully replaced by hydrogen bonds to solvent molecules after switching to the MLP. Therefore, switches starting from MM conformations with hydrogen bonds to a solvent molecule require less work. Furthermore, the system does not switch back to hydrogen bonds with sulfur during the short time of the backward switch.

We estimated the binding free energy of the ligand 19G to the protein MCL1 as $\Delta G^\mathrm{ML(I)/MM}_\mathrm{bind} = -35.3~\si{kJ.mol^{-1}} \pm 1.8~\si{kJ.mol^{-1}}$\cite{Q4Bio-Paper1}, underestimating the binding by $1.9~\si{kJ.mol^{-1}}$ compared to the experimental estimate ($-37.3~\si{kJ.mol^{-1}} \pm 0.1~\si{kJ.mol^{-1}}$\cite{Friberg2012}). After generating MLP(II) through transfer learning, we carried out six NEQ switching simulations for the protein-ligand complex and the solvated ligand. We calculated the mean of the resulting 36 binding free energies $\langle \Delta G^\mathrm{ML(II)/MM}_\mathrm{bind}\rangle = -37.2~\si{kJ.mol^{-1}} \pm 1.0~\si{kJ.mol^{-1}}$, which closely matches the experimental estimate. However, we note here that the protonation state of the ligand is unclear since its carboxylic acid group is likely deprotonated in solution. Therefore, this close agreement with the experimental estimate may have benefited from error cancellation. We chose the neutral ligand to avoid charge compensation during the initial MM AFE calculation.

\section{Conclusions}
In this work, we have developed a two-level QM/QM/MM approach that can be exploited to refine QM/MM potential energy surfaces by allowing for highly accurate quantum chemical calculations in quantum cores, which are embedded into the large QM region of a QM/MM model. In this way, a higher accuracy for substructures can be fed into a machine learning representation of the potential energy surface. 
This is brought about by projection-based embedding and transfer learning of a machine learning potential. As a result, the large QM region of the initial QM/MM model, that is typically described by DFT or any other semi-empirical approach, can be locally improved by correlated coupled cluster calculations as demonstrated in this work. Depending on the nature of the electronic structure, also multi-configurational approaches are applicable and even quantum computation can step in as a source of high-accuracy quantum energies, as we will demonstrate in a forthcoming paper.\cite{Q4Bio-Umbrella}

We demonstrated the capabilities of this approach for binding free energies of protein-drug complexes, typically investigated with MM-only approaches. Our approach relies on an end-state correction through nonequilibrium switching of MM-based alchemical free energy simulations. The nonequilibrium switching calculations are made feasible by training MLPs on QM/MM energies and forces for the protein-ligand complex and the solvated ligand.
These QM/MM-based MLPs were then improved through transfer learning with QM/QM/MM energies.

We validated our approach using the well-studied protein-ligand complex of the inhibitor 19G binding to myeloid cell leukemia 1. The QM/QM/MM embedding provided a locally corrected description of the electronic structure, which had a small but significant effect on the relative energy distributions. In fact, the QM/QM/MM-based MLP provided a binding free energy for MCL1-19G in close agreement with the experimental estimate, improving on our previous QM/MM-based approach\cite{Q4Bio-Paper1}.

Furthermore, we investigated the importance of energy derivatives for learning the potential energy surface with transfer learning from QM/QM/MM data. To this end, we trained a machine learning potential to predict the MM energies and forces. We then compared the performance of transfer learning from this MM-ML potential to QM/MM using only energies and using energies and forces. Training with only energies increased the errors for the forces, suggesting that significantly more data points are required to reach a similar accuracy as the ML potential trained with energies and forces. This demonstrates that energy derivatives are key for the efficient training of MLPs and provide additional incentive to develop analytical expressions for energy derivatives of highly accurate electronic structure methods, such as DLPNO-CCSD(T$_0$), within embedding frameworks, which are, however, currently unavailable.

We also want to note that the specific selection of the quantum cores may have a nonnegligible effect on the overall description. Very small embedded regions in Huzinaga-based embedding can decrease the accuracy of relative energies compared to a full DFT description since error cancellation intrinsic to DFT is lost\cite{Bensberg2020a}. However, these problems can be addressed by more accurate exchange-correlation functionals, such as double hybrid functionals, or by increasing the size of the quantum cores (for instance, through automated selection procedures\cite{Bensberg2019a}).

In future work, we will investigate active learning strategies for transfer learning to lower the computational cost of our approach. Such an active learning strategy aims to identify the structures for which the QM/QM/MM and QM/MM potential energy surfaces are significantly different, and a QM/QM/MM calculation is required to correct the PES locally.

\section*{Acknowledgments}

This work is part of the research project ``Molecular Recognition from Quantum Computing'' and is supported by Wellcome Leap as part of the Quantum for Bio (Q4Bio) program.
Moreover, we are grateful to Novo Nordisk for financial support through the Quantum for Life center in Copenhagen/Zurich, NNF20OC0059939.
This work was also created as part of NCCR Catalysis (grant number 180544), a National Centre of
Competence in Research funded by the Swiss National Science Foundation.
M.E.\ gratefully acknowledges an ETH Zurich Postdoctoral Fellowship. 
Co-funded by the European Union (ERC, DynaPLIX, SyG-2022 101071843, to K.L.-L.). Views and opinions expressed are however those of the authors only and do not necessarily reflect those of the European Union or the European Research Council. Neither the European Union nor the granting authority can be held responsible for them.
G.C.S. and W.B.J. acknowledge funding from the European Research Council (ERC) under the European Union’s Horizon 2020 research and innovation programme (grant agreement No 865870). G.C.S. and M.S.T. acknowledge funding from the Novo Nordisk Foundation, Grant number NNF22SA0081175, NNF Quantum Computing Programme and NNF20OC0060019, SolidQ. A.K. acknowledges support from the Novo Nordisk Foundation NNF20OC0062606 and NNF20OC0063268.

\section*{Data Availability}
The transfer learned machine learning potentials, the databases containing all QM/QM/MM and QM/MM energies, and the final work distributions for all NEQ simulations are available on Zenodo\cite{zenodo-archive}.

\section*{Author Contributions}

Moritz Bensberg: Methodology, Software, Investigation, Validation, Writing – original draft.

Marco Eckhoff: Methodology, Software, Investigation, Validation, Writing – original draft.

Raphael T. Husistein: Investigation, Validation, Software Writing – original draft.

Matthew S. Teynor: Software, Writing - Review \& Editing.

Valentina Sora:  Software, Writing - Review \& Editing.

William Bro-Jørgensen: Software, Writing - Review \& Editing.

F. Emil Thomasen: Software, Writing - Review \& Editing.

Anders Krogh: Conceptualization, Supervision, Writing - Review \& Editing.

Kresten Lindorff-Larsen:  Conceptualization, Supervision, Writing - Review \& Editing.

Gemma C. Solomon:  Conceptualization, Supervision, Writing - Review \& Editing.

Thomas Weymuth: Validation, Software, Supervision, Writing - Review \& Editing.

Markus Reiher: Conceptualization, Supervision, Writing - Review \& Editing.

\providecommand{\latin}[1]{#1}
\makeatletter
\providecommand{\doi}
  {\begingroup\let\do\@makeother\dospecials
  \catcode`\{=1 \catcode`\}=2 \doi@aux}
\providecommand{\doi@aux}[1]{\endgroup\texttt{#1}}
\makeatother
\providecommand*\mcitethebibliography{\thebibliography}
\csname @ifundefined\endcsname{endmcitethebibliography}
  {\let\endmcitethebibliography\endthebibliography}{}


\end{document}